\documentclass[11pt]{article}
\usepackage{amsfonts}
\newcommand{\eq}{\begin{equation}}
\newcommand{\en}{\end{equation}}
\newcommand{\eqn}{\begin{eqnarray}}
\newcommand{\enn}{\end{eqnarray}}

\newcommand{\beq}{\begin{equation}}
\newcommand{\eeq}{\end{equation}}
\newcommand{\tn}{\ensuremath{\tilde{n}}}
\newcommand{\ta}{\ensuremath{\tilde{a}}}
\newcommand{\tb}{\ensuremath{\tilde{b}}}
\newcommand{\tc}{\ensuremath{\tilde{c}}}
\newcommand{\tx}{\ensuremath{\tilde{x}}}
\newcommand{\ty}{\ensuremath{\tilde{y}}}
\newcommand{\ti}{\ensuremath{\tilde{I}}}
\newcommand{\tj}{\ensuremath{\tilde{J}}}
\newcommand{\tk}{\ensuremath{\tilde{K}}}

\newcommand{\pe}{\ensuremath{\varphi^{1}}}
\newcommand{\pz}{\ensuremath{\varphi^{2}}}
\newcommand{\pd}{\ensuremath{\varphi^{3}}}
\newcommand{\np}{\ensuremath{\|\varphi\|}}
\newcommand  {\Rbar} {{\mbox{\rm$\mbox{I}\!\mbox{R}$}}}

\begin{document}
\begin{titlepage}
\begin{flushright}
  PSU-TH-227\\
 CERN-TH/2000-068\\
\end{flushright}
\vspace{0.2cm}
\begin{center}
\begin{LARGE}
\textbf{The Vacua of 5d,
$\mathcal{N}=2$ Gauged Yang-Mills/Einstein/Tensor Supergravity:\\
Abelian Case}\footnote{ Work supported in part by the National
Science Foundation under Grant Number PHY-9802510.}
\end{LARGE}\\
\vspace{1.0cm}
\begin{large}
M. G\"{u}naydin$^{\dagger\ddagger}$ \footnote{murat@phys.psu.edu} and
M. Zagermann$^{\ddagger}$ \end{large}\footnote{zagerman@phys.psu.edu}  \\
\vspace{.35cm}
$^{\dagger}$ \emph{CERN, Theory Division \\
1211 Geneva 23, Switzerland} \\
\vspace{.3cm}
and \\
\vspace{.3cm}
$^{\ddagger}$ \emph{Physics Department \\
Penn State University\\
University Park, PA 16802, USA} \\
\vspace{0.5cm}
{\bf Abstract}
\end{center}
\begin{small}

We give a detailed study of the critical points of the potentials of the
simplest non-trivial $\mathcal{N}=2$ gauged Yang-Mills/Einstein supergravity theories
with tensor multiplets. The scalar field target space of these examples
is $SO(1,1)\times
SO(2,1)/SO(2)$. The possible gauge groups are $SO(2)\times U(1)_R$
and $SO(1,1)\times U(1)_R$, where $U(1)_{R}$ is a subgroup of the
R-symmetry group $SU(2)_{R}$, and $SO(2)$ and $SO(1,1)$ are subgroups of
the isometry group of the scalar manifold. The scalar potentials
of these theories consist of a contribution from the $U(1)_{R}$ gauging
and a contribution that is due to the presence of the
tensor fields.
We find that the latter contribution can change the form of
the supersymmetric extrema from maxima to saddle points. In addition, it
leads
 to novel critical points not present
in the corresponding gauged Yang-Mills/Einstein supergravity theories
\emph{without} the tensor multiplets.
For the $SO(2)\times U(1)_R$ gauged theory
these novel critical points correspond to anti-de
Sitter ground states. For the non-compact $SO(1,1)\times U(1)_R$ gauging,
the novel
ground states are de Sitter. The analysis of the critical points of
the potential carries over in a straightforward manner to the generic
 family of $\mathcal{N}=2$ gauged Yang-Mills/Einstein supergravity theories
with tensor multiplets whose scalar manifolds are of the form
$SO(1,1)\times SO(n-1,1)/SO(n-1)$.
\end{small}

\end{titlepage}

\renewcommand{\theequation}{\arabic{section}.\arabic{equation}}
\section{Introduction}
\setcounter{equation}{0}

In the last few years there has been a renewed intense interest
in  gauged
supergravity theories. The work
on AdS/CFT (anti-de Sitter/conformal field theory)  dualities in recent years
has reaffirmed the importance of gauged supergravity theories in various
dimensions to the understanding of the dynamics of M/superstring-theory \cite{jm,
GKP,EW98,earlier,agmoo}.    The  best studied example of this duality
is between the
  IIB superstring theory on the background manifold
$AdS_{5}\times S^{5}$ with $N$ units of five-form flux through the five-sphere
and   $4d$, $\mathcal{N}=4$ super
 Yang Mills theory with gauge
group $SU(N)$, which is a conformally invariant quantum field theory.
In the limit of small string coupling and large $N$, the classical
(i.e. tree level) IIB supergravity approximation becomes valid.
The lowest lying Kaluza Klein modes of IIB supergravity on
$AdS_{5}\times S^{5}$  are believed to form a consistent nonlinear
truncation\footnote{The consistency of the nonlinear
truncation for a subsector of the scalar manifold has been shown recently \cite
{LPT}.} \cite{GM,krv} which is described by five-dimensional $\mathcal{N}=8$
gauged supergravity\cite{GRW0,GRW1,PPvN}. Many aspects of the AdS/CFT
 correspondence, like eg. RG flows \cite{FGPW,npw},  can therefore
 be studied entirely within the framework of $5d$ gauged supergravity
due to the lack of interference with the higher Kaluza-Klein modes.

On the other hand, five-dimensional, $\mathcal{N}=2$ gauged
 supergravity theories naturally occur as effective field theories in
certain brane world scenarios based on heterotic
 M-theory compactifications \cite{HW1,EW2,LOSW1,ELPP}.
Since gauged supergravity theories typically also
allow for AdS ground states, they have recently been discussed
as a potential framework   for embedding the Randall/Sundrum
scenario \cite{RS} into M/string theory.

Several attempts in this direction have been made  (see
  eg. \cite{BC1,KLS,RK,KL,BC2,CLP}.)
Many of them focused on what we will later call
$\mathcal{N}=2$ ``gauged Maxwell/Einstein theories'' \cite{BC1,KLS,RK,KL,BC2}.
 It was found, however,
that the scalar potentials of these theories are not of the right form
to admit a supersymmetric embedding of Randall/Sundrum-type models \cite{RK,KL,BC2}.
The question whether this is a generic feature of all gauged supergravity
 theories provides one of the motivations to
study the potentials of more general gauged
supergravity theories in five dimensions.

Recently, we have
constructed  the  general gaugings
of $5d$,  $\mathcal{N}=2$ supergravity
coupled to vector as well as {\it tensor} multiplets
\cite{gz99}. This was an extension and generalization of  earlier
work on the gaugings of $\mathcal{N}=2$ supergravity coupled to
vector multiplets only
\cite{GST1,GST2,GST3,GST4,GST5}.

Starting point of our construction were the ungauged Maxwell/Einstein
supergravity theories (MESGT's) of ref. \cite{GST1}, which describe the
coupling of Abelian vector multiplets to supergravity.
These theories have a global symmetry group of the form $SU(2)_{R}\times G$, where
$G$ is the subgroup of the isometry group of the scalar field target manifold
that extends to a global symmetry group of the full Lagrangian, and $SU(2)_R$
denotes   the R-symmetry group of the $\mathcal{N}=2$ supersymmetry algebra.
In general, there are various ways to turn a subgroup of
 $SU(2)_{R}\times G$ into a local gauge group. We will use different names
for these different possibilities \cite{GST2,gz99}:

We refer to theories in which $U(1)_{R}\subset SU(2)_{R}$ is gauged
 as ``gauged Maxwell/Einstein supergravity theories''.

In order to gauge a subgroup $K$
of $G$, a subset of the vector fields of the ungauged theory has to transform
in the adjoint representation of $K$. If such a group $K$ exists,
there are two possibilities:\\
(i) There are additional vector fields
outside the adjoint of $K$ which transform nontrivially under $K$.
These vector fields have to be dualized to ``self-dual'' antisymmetric
tensor fields
in order to perform the gauging of $K$ in a supersymmetric way \cite{gz99}.
\footnote{ We should note that the gauging of $\mathcal{N}=8$ Poincar\'{e}
supergravity in $5d$ requires the dualization of twelve of the vector
fields of the $\mathcal{N}=8$ Poincar\'{e} supermultiplet
to self-dual tensor fields \cite{GRW0,GRW1,PPvN}
for completely analogous reasons.}
\\
(ii) If there are no vector fields outside the adjoint of $K$,
or if the additional vectors are all singlets under $K$ (``spectator vector fields''),
 the gauging of $K$ proceeds in a straightforward way, and no tensor fields have
 to be introduced \cite{GST2}.\\
\indent In order to distinguish between gaugings of $U(1)_{R}$ and $K$, we will
refer to theories in which $K$ is gauged as
``Yang-Mills/Einstein supergravity theories'' (``with or without tensor fields'',
 depending on which of the possibilities (i) or (ii)
 is realized)\footnote{We will use the term ``Yang-Mills'' also when $K$ is
  Abelian (as  is the case for our examples in Section 4).}

The most general gauging in this framework is then obviously a simultaneous
 gauging of $U(1)_{R}$ and $K$. For consistency with our terminology, we
will sometimes use the term ``gauged Yang-Mills/Einstein supergravity theories
(with or without tensor multiplets)''
for this type of gauging.

As for the scalar potentials that are introduced by  these different types
of gaugings, one makes the following observation\cite{GST2,gz99}:\\
(i) The gauging of $U(1)_{R}$ introduces a scalar potential, which in all known cases\\
\indent a) either has a maximum that corresponds to an anti-de Sitter space
or\\
\indent b) vanishes identically or\\
\indent c) has no critical points at all.\\
(ii) The gauging of $K$ introduces no potential when no vector fields have to
be dualized to tensor fields.\\
(iii) If tensor fields have to be introduced, the gauging of $K$ introduces
a scalar potential which is positive semidefinite and can therefore not
lead to AdS vacua.\\
(iv) The simultaneous gauging of $U(1)_{R}$ and $K$ leads to a scalar potential which
 is simply the sum of the potentials that would result from the gaugings
of $U(1)_{R}$ and $K$ alone. The critical points of this combined potential
have not yet been fully investigated.\\
\indent The purpose of this paper is to give
an explicit example of a
gauged Yang-Mills/Einstein supergravity theory with tensor fields
which is simple enough to admit a complete analysis of its scalar potential.
The  model we discuss describes the coupling of one vector multiplet
and one self-dual tensor multiplet (which contains two real tensor fields) to
 supergravity. The three scalar fields
from the vector/tensor multiplets parametrize the
space $\mathcal{M}=SO(1,1)\times SO(2,1)/SO(2)$, and
the possible gauge groups are
$U(1)_{R}\times SO(2)$ and $U(1)_{R}\times SO(1,1)$.
We will find that the structure of the resulting scalar potentials is much
 richer than for gaugings without tensor fields.

The organization of the paper is as follows. Section 2 briefly summarizes the
most general form of a gauged Yang-Mills/Einstein supergravity theory
with tensor fields. Section 3
discusses some general properties of the scalar potentials of these
theories. The ungauged
MESGT with scalar manifold $\mathcal{M}=SO(1,1)\times SO(2,1)/SO(2)$,
its $U(1)_{R}\times SO(2)$ and $U(1)_{R}\times SO(1,1)$ gaugings
and the resulting scalar potentials
are analyzed in section 4, which represents the main part of this paper.
Section 5 discusses the generalization to the scalar manifolds
$SO(1,1)\times SO(n-1,1)/SO(n-1)$, and Section 6 finally ends with some conclusions.
 An appendix summarizes the ``very special geometry'' of the
ungauged
$\mathcal{M}=SO(1,1)\times SO(2,1)/SO(2)$ theory.

\renewcommand{\theequation}{\arabic{section}.\arabic{equation}}
\section{Gauged Yang-Mills/Einstein supergravity with tensor
 fields}
\setcounter{equation}{0}
In this section, we briefly review the most  relevant features
of $\mathcal{N}=2$ gauged
Yang-Mills/Einstein supergravity theories coupled to tensor multiplets
\cite{gz99}.
Unless otherwise stated, our conventions will coincide
with those of ref.
\cite{GST1,GST2,gz99}, where further details can be found.
In particular, we will use the  metric signature
$(-++++)$ and impose the `symplectic' Majorana condition on
all fermionic quantities.

The fields of the $\mathcal{N} =2$ supergravity multiplet are the
f\"{u}nfbein $e_{\mu}^{m}$, two gravitini $\Psi_{\mu}^{i}$ ($i=1,2$)
and a vector field $A_{\mu}$. An $\mathcal{N} =2$ vector multiplet contains
a vector field $A_{\mu}$, two spin-$1/2$ fermions $\lambda^{i}$ and one real
scalar field $\varphi$. The fermions of each of these multiplets transform as
doublets under the $USp(2)_{R}\cong SU(2)_{R}$ R-symmetry
group of the
$\mathcal{N} =2$ Poincar\'{e} superalgebra; all other fields are
$SU(2)_{R}$-inert. A tensor field satisfying a 5-dimensional ``self-duality''
 condition
must necessarily be complex \cite{PTvN}. We choose to work with the
real and imaginary parts of the complex tensors. A
self-dual
$\mathcal{N}=2$ tensor multiplet contains such a pair of  tensor fields, four
spin-$1/2$ fermions (i.e. two
$SU(2)_{R}$ doublets) and two scalars.

The general coupling of $m$ self-dual tensor multiplets to
$\mathcal{N}=2$ gauged Yang-Mills/Einstein supergravity was given
in \cite{gz99}. The field content of
these theories is
\begin{equation}
\{ e_{\mu}^{m}, \Psi_{\mu}^{i}, A_{\mu}^{I}, B_{\mu\nu}^{M},
\lambda^{i\ta}, \varphi^{\tx}\}
\end{equation}
where
\begin{eqnarray*}
I&=& 0,1, \ldots n \\
M&=& 1,2, \ldots 2m \\
\ta&=& 1,\ldots, \tn\\
\tx&=& 1,\ldots, \tn,
\end{eqnarray*}
with $\tn=n+2m$.
Note that we have combined the `graviphoton' with the $n$ vector fields
of the $n$ vector multiplets into a single $(n+1)$-plet of vector fields
$A_{\mu}^{I}$ labelled by the index $I$. Also, the spinor and scalar fields
of the vector and tensor multiplets are combined into
$\tn$-tupels of spinor and scalar fields.
The indices $\ta, \tb, \ldots$ and $\tx, \ty, \ldots$ are the
flat and curved indices, respectively, of
the $\tn$-dimensional target  manifold $\mathcal{M}$ of the scalar fields.
The metric, vielbein and spin connection on $\mathcal{M}$ will be
denoted by $g_{\tx\ty}$, $f_{\tx}^{\ta}$ and $\Omega_{\tx}^{\ta\tb}$,
respectively. The $SU(2)_{R}$ index $i$ is raised and lowered
with the antisymmetric metric $\varepsilon _{12}=\varepsilon ^{12}=1$
according to
\begin{displaymath}
  X^i=\varepsilon ^{ij}X_j\,,\qquad X_i=X^j\varepsilon _{ji}\,.
\end{displaymath}
The fermions
$\Psi_{\mu}^{i}$ and $\lambda^{i\ta}$ are $U(1)_{R}$-charged, whereas the fields
$\varphi^{\tx}$, $\lambda^{i\ta}$ and $B_{\mu\nu}^{M}$ carry charge under $K$.

Denoting the $U(1)_{R}$ and $K$ coupling constants by $g_{R}$ and $g$,
respectively, the $(U(1)_{R}\times K)$ gauge covariant derivatives of
these fields are
as follows ($\nabla$ denotes the ordinary spacetime covariant derivative)
\begin{eqnarray}
\mathfrak{D}_{\mu}\Psi_{\nu}^{i}
&\equiv& \nabla_{\mu}\Psi_{\nu}^{i}+g_{R}V_{I}A_{\mu}^{I}\delta^{ij}
\Psi_{\nu j}\nonumber\\
 \mathfrak{D}_{\mu}
\lambda^{i\ta}
&\equiv&  \nabla_{\mu}\lambda^{i\ta}+g_{R}V_{I}A_{\mu}^{I}\delta^{ij}
\lambda_{j}^{\ta}  +gA_{\mu}^{I} L_{I}^{\ta\tb}\lambda^{i\tb}  \nonumber\\
\mathfrak{D}_{\mu}
\varphi^{\tx}&\equiv&
\partial_{\mu}\varphi^{\tx}+gA_{\mu}^{I}K_{I}^{\tx}\nonumber\\
\mathfrak{D}_{\mu}B_{\nu\rho}^{M}&\equiv& \nabla_{\mu} B_{\nu\rho}^{M}
+gA_{\mu}^{I}\Lambda_{IN}^{M}B_{\nu\rho}^{N}.
\end{eqnarray}
Here, $K_{I}^{\tx}$ are the Killing vector fields on $\mathcal{M}$ that generate
 the subgroup $K$ of its isometry group. The $\varphi$-dependent
matrices $L_{I}^{\ta\tb}$ and the \emph{constant} matrices $\Lambda_{IN}^{M}$ are the
 $K$-transformation matrices
 of $\lambda^{i\ta}$ and $B_{\mu\nu}^{M}$, respectively.
The $V_I$ are some  constants that define
the  linear combination of the vector fields
$A_{\mu}^{I}$ that is used as  the $U(1)_R$-gauge field
\begin{equation}\label{AU1}
A_{\mu}{[}U(1)_R{]} = V_{I}A_{\mu}^{I}.
\end{equation}
They have to be constrained by
\begin{equation}\label{Vf}
V_{I}f_{JK}^{I}=0,
\end{equation}
with $f_{JK}^{I}$ being the structure constants of $K$ \footnote{If there are
 spectator vector fields among the $A_{\mu}^{I}$, the corresponding $f_{IJ}^{K}$
are just zero.}

We denote the curls  of the vector fields $A_{\mu}^{I}$ by $F_{\mu\nu}^{I}$.
The non-Abelian field strengths $\mathcal{F}_{\mu\nu}^{I}\equiv
F_{\mu\nu}^{I}+gf_{JK}^{I}A_{\mu}^{J}A_{\nu}^{K}$ ($I=0,1,\ldots n$)
of the gauge group $K$ and the self-dual tensor fields $B_{\mu\nu}^M$
($M=1,2\ldots,2m$)  are grouped together to define the tensorial
quantity $\mathcal{H}_{\mu\nu}^{\ti}=(\mathcal{F}_{\mu\nu}^I,B_{\mu\nu}^M)$
 with $\ti=0,1,\ldots, n+2m$.

The Lagrangian is then given by (up to 4-fermion terms) \cite{gz99}

\begin{eqnarray}\label{Lagrange2}
e^{-1}\mathcal{L}&=& -\frac{1}{2}R(\omega)-\frac{1}{2}
{\bar{\Psi}}_{\mu}^{i}\Gamma^{\mu\nu\rho}\mathfrak{D}_{\nu}\Psi_{\rho i}-
\frac{1}{4}{\stackrel{\scriptscriptstyle{o}}{a}}_{\ti\tj}\mathcal{H}_{\mu\nu}^{\ti}
\mathcal{H}^{\tj\mu\nu}
\nonumber\cr
& & -\frac{1}{2}{\bar{\lambda}}^{i\ta}\left(\Gamma^{\mu}\mathfrak{D}_{\mu}
\delta^{\ta\tb}+
\Omega_{\tx}^{\ta\tb}\Gamma^{\mu}\mathfrak{D}_{\mu}\varphi^{\tx}\right)
\lambda_{i}^{\tb}-
\frac{1}{2}g_{\tx\ty}(\mathfrak{D}_{\mu}\varphi^{\tx})(\mathfrak{D}^{\mu}
\varphi^{\ty})\nonumber\cr
&& -\frac{i}{2}{\bar{\lambda}}^{i\ta}\Gamma^{\mu}\Gamma^{\nu}\Psi_{\mu i}
f_{\tx}^{\ta}\mathfrak{D}_{\nu}\varphi^{\tx}+ \frac{1}{4}h_{\ti}^{\ta}
{\bar{\lambda}}^{i\ta}\Gamma^{\mu}\Gamma^{\lambda\rho}\Psi_{\mu i}
\mathcal{H}_{\lambda\rho}^{\ti}
\nonumber\cr
&&+\frac{i}{2\sqrt{6}}\left(\frac{1}{4}\delta_{\ta\tb}h_{\ti}+T_{\ta\tb\tc}
h_{\ti}^{\tc}\right)
{\bar{\lambda}}^{i\ta}\Gamma^{\mu\nu}\lambda_{i}^{\tb}
\mathcal{H}_{\mu\nu}^{\ti}\nonumber\cr
&& -\frac{3i}{8\sqrt{6}}h_{\ti}\left[{\bar{\Psi}}_{\mu}^{i}
\Gamma^{\mu\nu\rho\sigma}
\Psi_{\nu i}\mathcal{H}_{\rho\sigma}^{\ti}+2{\bar{\Psi}}^{\mu i}
\Psi_{i}^{\nu}\mathcal{H}_{\mu\nu}^{\ti}\right]\nonumber\cr
&& + \frac{e^{-1}}{6\sqrt{6}}C_{IJK}\varepsilon^{\mu\nu\rho\sigma\lambda}
\left\{ F_{\mu\nu}^{I}F_{\rho\sigma}^{J}A_{\lambda}^{K} + \frac{3}{2}g
F_{\mu\nu}^{I}A_{\rho}^{J}(f_{LF}^{K}A_{\sigma}^{L}A_{\lambda}^{F})\right.
\nonumber\cr
&& \qquad\qquad\qquad\qquad+\left.
\frac{3}{5}g^{2}(f_{GH}^{J}A_{\nu}^{G}A_{\rho}^{H})
(f_{LF}^{K}A_{\sigma}^{L}A_{\lambda}^{F})A_{\mu}^{I}\right\}\nonumber\cr
&&+\frac{e^{-1}}{4g}\varepsilon^{\mu\nu\rho\sigma\lambda}\Omega_{MN}
B_{\mu\nu}^{M}\mathfrak{D}_{\rho}B_{\sigma\lambda}^{N}\nonumber\\
&&+g{\bar{\lambda}}^{i\ta}\Gamma^{\mu}\Psi_{\mu i}W_{\ta}+
g{\bar{\lambda}}^{i\ta}\lambda_{i}^{\tb}W_{\ta\tb}-g^{2}P \nonumber\\
 &&-\frac{i\sqrt{6}}{8} g_{R} {\bar{\Psi}}_{\mu}^{i}
\Gamma^{\mu\nu}
\Psi_{\nu}^{j}\delta_{ij} P_{0}-\frac{1}{\sqrt{2}}
g_{R}{\bar{\lambda}}^{i\ta}
\Gamma^{\mu}
\Psi_{\mu}^{j}\delta_{ij} P_{\ta}\nonumber\\
&&+\frac{i}{2\sqrt{6}}g_{R}{\bar{\lambda}}^{i\ta}
\lambda^{j\tb}\delta_{ij} P_{\ta\tb}-g_{R}^{2}P^{(R)}
.
\end{eqnarray}
The transformation laws are (to leading order in fermion fields)
\begin{eqnarray}\label{trafo2}
\delta e_{\mu}^{m}&=& \frac{1}{2}{\bar{\varepsilon}}^{i}
\Gamma^{m}\Psi_{\mu i}\nonumber\cr
\delta \Psi_{\mu}^{i} &=&\mathfrak{D}_{\mu}\varepsilon^{i}+\frac{i
}
{4\sqrt{6}}h_{\ti}(\Gamma_{\mu}^{\:\:\:\nu\rho}-4\delta_{\mu}^{\nu}
\Gamma^{\rho})\mathcal{H}_{\nu\rho}^{\ti}\varepsilon^{i}
+\frac{i}{2\sqrt{6}}g_{R}P_{0}\Gamma_{\mu}
\delta^{ij}\varepsilon_{j}\nonumber\cr
\delta A_{\mu}^{I}&=& \vartheta_{\mu}^{I}\nonumber\cr
\delta B_{\mu\nu}^{M} &=& 2\mathfrak{D}_{[\mu}\vartheta_{\nu]}^{M} +
\frac{\sqrt{6}g}{4}
\Omega^{MN}h_{N}{\bar{\Psi}}^{i}_{[\mu}\Gamma_{\nu]}\varepsilon_{i}
+\frac{ig}{4}\Omega^{MN}h_{N\ta}{\bar{\lambda}}^{i\ta}\Gamma_{\mu\nu}
\varepsilon_{i}\nonumber\\
\delta \lambda^{i\ta}  &=& -\frac{i}{2}f_{\tx}^{\ta}
\Gamma^{\mu}(\mathfrak{D}_{\mu}
\varphi^{\tx})\varepsilon^{i} + \frac{1}{4}h_{\ti}^{\ta}
\Gamma^{\mu\nu}
\varepsilon^{i}\mathcal{H}_{\mu\nu}^{\ti}+gW^{\ta}\varepsilon^{i}
+\frac{1}{\sqrt{2}}g_{R}P^{\ta}\delta^{ij}\varepsilon_{j} \nonumber \\
\delta \varphi^{\tx}&=&\frac{i}{2}f^{\tx}_{\ta}{\bar{\varepsilon}}^{i}
\lambda_{i}^{\ta}
\end{eqnarray}
with
\begin{equation}
\vartheta_{\mu}^{\ti}\equiv -\frac{1}{2}h_{\ta}^{\ti}{\bar{\varepsilon}}^{i}
\Gamma_{\mu}\lambda_{i}^{\ta}+\frac{i\sqrt{6}}{4}h^{\ti}
{\bar{\Psi}}_{\mu}^{i}\varepsilon_{i}.
\end{equation}

The various scalar field dependent quantities
 $\stackrel{\scriptscriptstyle{o}}{a}_{\ti\tj}$, $h_{\ti}$, $h^{\ti}$, $h_{\ti}^{\ta}$,
  $h^{\ti\ta}$ and
$T_{\ta\tb\tc}$ that contract the different types of indices are already present
 in the corresponding \emph{ungauged} MESGT's and describe the
  ``very special''geometry of the scalar manifold $\mathcal{M}$ (see \cite{GST1}
  for details).
The ungauged MESGT's also contain a constant symmetric tensor $C_{\ti\tj\tk}$.
If the gauging of $K$ involves the introduction of tensor fields, the coefficients
of the type $C_{MNP}$ and $C_{IJM}$ have to vanish \cite{gz99}. The only
components that survive such a  gauging are thus $C_{IJK}$, which appear in the
Chern-Simons-like  term of (\ref{Lagrange2}), and $C_{IMN}$, which
are related to  the transformation matrices of the tensor fields by
\begin{displaymath}
\Lambda_{IN}^{M}=\frac{2}{\sqrt{6}}\Omega^{MP}C_{IPN}.
\end{displaymath}
Here $\Omega^{MN}$ is the inverse of  $\Omega_{MN}$, which is a (constant) invariant
antisymmetric tensor of the gauge
group $K$:
\begin{equation}
\Omega_{MN}=-\Omega_{NM}, \qquad \Omega_{MN}\Omega^{NP}=\delta_{M}^{P}.
\end{equation}

The terms proportional to
\begin{eqnarray}
W^{\ta}(\varphi)&=&-\frac{\sqrt{6}}{8}h_{M}^{\ta}\Omega^{MN}h_{N}\nonumber\\
W^{\ta\tb}(\varphi)&=&-W^{\tb\ta}(\varphi)
= ih_{\, }^{J[\ta}K_{J}^{\tb]}+\frac{i\sqrt{6}}{4}
h^{J}K_{J}^{[\ta;\tb]}
\end{eqnarray}
(the semicolon denotes covariant differentiation on the target space $\mathcal{M}$)
and the potential  term
\begin{equation}\label{P}
P(\varphi)=2W_{\ta}W^{\ta}
\end{equation}
are due to the presence of the tensor fields.

The supersymmetric gauging of the $U(1)_{R}$-factor, on the other hand,
 introduces the terms proportional to
\begin{eqnarray}\label{U1cons}
P^{\ta}(\varphi)&=&\sqrt{2}h^{\ta I}V_{I}\label{U1cons1}\\
P_{0}(\varphi)&=&2 h^{I}V_{I}\label{U1cons2}\\
P_{\ta\tb}(\varphi)&=&\frac{1}{2}\delta_{\ta\tb}P_{0}+2\sqrt{2}
T_{\ta\tb\tc}P^{\tc}\label{U1cons3}
\end{eqnarray}
in (\ref{Lagrange2}) and (\ref{trafo2}) and leads to the scalar potential
contribution
\begin{equation}
P^{(R)}(\varphi)=-(P_{0})^{2}+P_{\ta}P^{\ta}.
\end{equation}

\section{Some general properties of the scalar potential}
\setcounter{equation}{0}
As summarized in the previous section, the simultaneous gauging of
 $U(1)_{R}\subset SU(2)_{R}$ and a subgroup
$K\subset G$ of the isometry group $G$ of the vector/tensor
 multiplets moduli space
$\mathcal{M}$  leads to a scalar potential of the form

\begin{equation}
e^{-1}\mathcal{L}_{pot}= -g^{2}P-g_{R}^{2}P^{(R)},
\end{equation}
where $P^{(R)}$ arises from the gauging of $U(1)_{R}$, whereas $P$ is nonzero
if and only if some $K$-charged vector fields $A_{\mu}^{M}$ had to be
 dualized to tensor fields $B_{\mu\nu}^{M}$ in order to perform
 the gauging of $K$ in a supersymmetric way.
In the remainder we will write
\begin{equation}
P_{tot}:=P+\lambda P^{(R)}, \qquad \textrm{ with }
\lambda:=\frac{g_{R}^{2}}{g^{2}}
\end{equation}
so that
\begin{equation}
e^{-1}\mathcal{L}_{pot}=-g^{2}P_{tot}.
\end{equation}
The potentials $P$ and $P^{(R)}$ are given by
\begin{eqnarray}
P&=&2W_{\ta}W^{\ta} \\ \nonumber
P^{(R)}&=&-(P_{0})^{2}+P_{\ta}P^{\ta}. \nonumber
\end{eqnarray}

Using $h^{\ti}_{\ta}h^{\ta}_{\tj}=\delta_{\tj}^{\ti}-h^{\ti}h_{\tj}$ \cite{GST1},
 it is easy to verify that $W^{\ta}$ and $P^{\ta}$ are orthogonal:
\begin{displaymath}
W_{\ta}P^{\ta}=0.
\end{displaymath}
Contracting $\langle \delta \lambda^{i\ta}\rangle =0$ with $W^{\ta}$ and
$P^{\ta}$ then shows that an $\mathcal{N}=2$ supersymmetric ground state
requires
\begin{equation}
\langle W^{\ta}\rangle=\langle P^{\ta}\rangle=0.
\end{equation}
This implies, in particular, that the cosmological constant of
an $\mathcal{N}=2$ supersymmetric vacuum is given by $P^{(R)}(\varphi_{c})$
alone, i.e. $P(\varphi_{c,susy})=0$, as has also been
pointed out in \cite{KL}. Nevertheless, $P$ can still have a non-trivial effect
on the \emph{form} of a  supersymmetric critical point, i.e. it can change
 it from a maximum to a saddle point.
In addition, there might be critical points  which do not
preserve the full $\mathcal{N}=2$ supersymmetry and therefore \emph{can} have
$P(\varphi_{c})\neq 0$. We will see examples for all this in the
next section.

Using \cite{GST1}
\begin{displaymath}
C_{\ti\tj\tk}h^{\tk}=h_{\ti}h_{\tj}-\frac{1}{2}h_{\ti\ta}h_{\tj}^{\ta},
\end{displaymath}
$P$ and $P^{(R)}$ can be expressed in a more compact form which will
facilitate the analysis of the critical points:

\begin{eqnarray}
P&=&\frac{3\sqrt{6}}{16}h^{I}\Lambda_{I}^{MN}h_{M}h_{N}\label{PLambda}\\
P^{(R)}&=&-4C^{IJ\tk}V_{I}V_{J}h_{\tk},
\end{eqnarray}
where we have defined
\begin{eqnarray}
\Lambda_{I}^{MN}&\equiv&\Lambda_{IP}^{M}\Omega^{PN}=
\frac{2}{\sqrt{6}}\Omega^{MR}C_{IRP}\Omega^{PN}\\
C^{\ti\tj\tk}&\equiv&{\stackrel{\scriptscriptstyle{\circ}}{a}}^{\ti\ti'}
{\stackrel{\scriptscriptstyle{\circ}}{a}}^{\tj\tj'}
{\stackrel{\scriptscriptstyle{\circ}}{a}}^{\tk\tk'}C_{\ti'\tj'\tk'}
\end{eqnarray}
with ${\stackrel{\scriptscriptstyle{\circ}}{a}}^{\ti\tj}$ being the inverse of
 ${\stackrel{\scriptscriptstyle{\circ}}{a}}_{\ti\tj}$.

If $\mathcal{M}$ is associated with a  Jordan algebra
\footnote{We recall that the MESGT's associated with Jordan algebras
are those for which the cubic form defined by the symmetric tensor
$C_{\ti\tj\tk} $ can be identified with
 the norm form of a Jordan algebra of degree
three.}
 \cite{GST1},
one has (componentwise)
\begin{displaymath}
C^{\ti\tj\tk}=C_{\ti\tj\tk}=\textrm{const.}
\end{displaymath}
In this case, because of $C^{IJM}=C_{IJM}=0$, $P^{(R)}$ simplifies to
\begin{equation}\label{PRJ}
P^{(R)}=-4C^{IJK}V_{I}V_{J}h_{K}\qquad \textrm{(for the Jordan family)}
\end{equation}
with \emph{constant} $C^{IJK}=C_{IJK}$ and summation over $K$ instead of $\tk$.

The critical points of $P^{(R)}$ have been analyzed in \cite{GST2} for the
purely $U(1)_{R}$-gauged MESGTs of  the  Jordan type. It was found that they are
 characterized by the ``dual'' element
\begin{equation}\label{Vdual}
V^{\#\ti}\equiv\sqrt{\frac{2}{3}}C^{\ti\tj\tk}V_{\tj}V_{\tk}
\end{equation}
of $V_{\ti}$. Three cases could be distinguished:
\begin{enumerate}
\item $V^{\#\ti}=0$. In this case, the scalar potential $P^{(R)}$ vanishes identically,
 leading to  Minkowski ground states with broken supersymmetry.

\item $V^{\#\ti}$ is in the ``domain of positivity'' of the corresponding
Jordan algebra $J$. In this case, there exists precisely one critical point,
which sits at the unique global maximum of the scalar potential $P^{(R)}$
and corresponds to an Anti-de Sitter ground state with unbroken $\mathcal{N}=2$
supersymmetry and unbroken global Aut(J)-invariance, where Aut(J) denotes the
automorphism group of the Jordan algebra $J$.

\item $V^{\#\ti}$ is non-zero and not in the domain of positivity of $J$.
In this case, the scalar potential $P^{(R)}$ has no critical points at all.

\end{enumerate}

In order to get a better understanding as to whether and how the  presence
of the tensor field related potential
$P$ changes this picture, we will analyze the simplest non-trivial example
 of a gauged Yang-Mills/Einstein supergravity theory with tensor multiplets
in full detail in the next section.

\renewcommand{\theequation}{\arabic{section}.\arabic{equation}}
\section{The simplest nontrivial example: $\mathcal{M}=SO(1,1)\times SO(2,1)/SO(2)$}

\subsection{The ungauged theory}
\setcounter{equation}{0}
The \emph{ungauged} MESGT with the scalar manifold
$\mathcal{M}=SO(1,1)\times SO(2,1)/SO(2)$ allows the construction of two of
the simplest non-trivial examples of a
 gauged Yang-Mills/Einstein supergravity theory with tensor multiplets.
Let us consider this  \emph{ungauged} theory first. It belongs to the generic
Jordan family\footnote{The ``generic Jordan family'' consists of the
MESGT's with scalar manifolds of the form $\mathcal{M}=SO(1,1)\times SO(n-1,1)/SO(n-1)$.}
 and describes the coupling of three Abelian vector multiplets to
supergravity. Consequently, the field content is
\begin{equation}
\{ e_{\mu}^{m}, \Psi_{\mu}^{i}, A_{\mu}^{\ti}, \lambda^{i\ta}, \varphi^{\tx}\}
\end{equation}
with
\begin{eqnarray*}
i&=& 1,2\\
\ti&=& 0,1,\ldots, 3\\
\ta&=& 1,\ldots, 3\\
\tx&=& 1,\ldots, 3,
\end{eqnarray*}
 where the three scalar fields $\varphi^{\tx}$ parametrize the target space
$\mathcal{M}=SO(1,1)\times SO(2,1)/SO(2)$.
The latter can be described as the hypersurface
\begin{displaymath}
N(\xi)=\left( \frac{2}{3}\right)^{\frac{3}{2}}C_{\ti\tj\tk}\xi^{\ti}
\xi^{\tj}\xi^{\tk} = 1
\end{displaymath}
in a four-dimensional ambient vector space parametrized by coordinates $\xi^{\ti}$.
In the case at hand, this vector space can be identified with the
Jordan algebra
\begin{displaymath}
J=\Rbar \oplus \Sigma_{3},
\end{displaymath}
where $\Sigma_{3}$ is the Jordan algebra of degree two corresponding to a
 quadratic form $Q$ with signature $(+,-,-)$ \cite{GST1}. In the most natural
  basis of this Jordan algebra, $N(\xi)$
takes on the following form
\begin{displaymath}
N(\xi)=\sqrt{2} \xi^{0}\left[(\xi^{1})^{2}-(\xi^{2})^{2}   -(\xi^{3})^{2}
\right],
\end{displaymath}
where the normalization factor $\sqrt{2}$ ensures that the unique selfdual
point
$\xi^{\ti}=\xi^{\#\ti}$ (i.e. the ``basepoint'' $c^{\ti}$ of the
Jordan algebra \cite{GST1}) really lies on the hypersurface $N(\xi)=1$, or equivalently,
 that there is a point on $\mathcal{M}$
where ${\stackrel{\scriptscriptstyle{\circ}}{a}}_{\ti\tj}=\delta_{\ti\tj}$
 \cite{GST1}.\footnote{In our parametrization, $c^{\ti}=(\frac{1}{\sqrt{2}},1,0,0)$,
 which corresponds to $\varphi^{\tx}=(1,0,0)$  (cf.    the Appendix).}

Hence, the nonvanishing $C_{\ti\tj\tk}$ are
\begin{eqnarray}\label{CIJK}
C_{011}&=&\frac{\sqrt{3}}{2}\nonumber\\
C_{022}&=&C_{033}=-\frac{\sqrt{3}}{2}.
\end{eqnarray}
The constraint $N=1$ can be solved by
\begin{eqnarray}
\xi^{0}&=&\frac{1}{\sqrt{2}\|\varphi\|^{2}}\\
\xi^{1}&=& \varphi^{1}\\
\xi^{2}&=& \varphi^{2}\\
\xi^{3}&=& \varphi^{3}
\end{eqnarray}
with
\begin{displaymath}
\|\varphi\|^{2}\equiv (\varphi^{1})^{2}- (\varphi^{2})^{2} - (\varphi^{3})^{2}.
\end{displaymath}
Obviously, the hypersurface $N=1$ decomposes into three disconnected
components:\\
(i) $\np^{2}>0$ and $\pe>0$\\
(ii) $\np^{2}<0$ \\
(iii) $\np^{2}>0$ and $\pe<0$.

In the following, we will consider the ``positive timelike'' region (i)
only, since in region (ii), $g_{\tx\ty}$ and
 ${\stackrel{\scriptscriptstyle{\circ}}{a}}_{\ti\tj}$ are not positive definite
 (see the Appendix),    and  region (iii) is isomorphic to region (i).

All the scalar field dependent quantities in the Lagrangian and the
supersymmetry transformation laws can be derived from $N(\xi)$, and
they are listed in  the Appendix.
\subsection{The $U(1)_{R}\times SO(2)$ gauging}
We will now turn the above ungauged   $SO(1,1)\times SO(2,1)/SO(2)$ model
into a gauged Yang-Mills/Einstein supergravity theory with tensor fields.

The isometry group  of the scalar manifold $\mathcal{M}$ is $G=SO(2,1)\times SO(1,1)$,
 which is simply the invariance group of $N(\xi)$. There are now two different ways
  to construct a Yang-Mills/Einstein supergravity theory with  tensor multiplets:
   Either one gauges the
compact subgroup $SO(2)\subset SO(2,1)$, or one gauges the noncompact subgroup
 $SO(1,1)\subset SO(2,1)$. We will  focus on  the compact gauging first and  discuss the
noncompact $SO(1,1)$ gauging in  the next subsection.
 The $SO(2)$ subgroup of $SO(2,1)$ rotates $\xi^{2}$ and $\xi^{3}$ into each other
 and therefore acts nontrivially on the vector fields $A_{\mu}^{2}$ and $A_{\mu}^{3}$.
  Hence, gauging this $SO(2)$ requires the dualization of
$A_{\mu}^{2}$ and $A_{\mu}^{3}$ to antisymmetric tensor fields. Accordingly,
we decompose the index $\ti$ as follows
\begin{displaymath}
\ti=(I,M)
\end{displaymath}
with $I,J,K,\ldots=0,1$ and $M,N,P,\ldots=2,3$.

It is easy to verify  that our $C_{\ti\tj\tk}$ in eqs. (\ref{CIJK}) are consistent
with the requirements $C_{IJM}=C_{MNP}=0$ for this type of gauging.
Having a closer look at the $C_{\ti\tj\tk}$ of the type $C_{IMN}$
we also see     that $C_{1MN}$ is zero, whereas $C_{0MN}$ is non-vanishing.
This means, because of $\Lambda_{IN}^{M}\sim \Omega^{MP}C_{IPN}$, that the vector
 field $A_{\mu}^{0}$ plays the r\^{o}le of the $SO(2)$-gauge field, whereas
 $A_{\mu}^{1}$ is just a ``spectator vector field'' with respect to  the
$SO(2)$-gauging.

In addition to this $SO(2)$-gauging, one can now use
 a linear combination $A_{\mu}[U(1)_{R}]=A_{\mu}^{I}V_{I}$ of the
vector fields $A_{\mu}^{0}$ and $A_{\mu}^{1}$ as the $U(1)_{R}$-gauge field,
and simultaneously gauge $U(1)_{R}$ and $SO(2)$. The result is a \emph{gauged}
 Yang-Mills/Einstein supergravity theory with tensor fields with the full gauge
 group $U(1)_{R}\times SO(2)$.

In our parametrization, the resulting potentials $P$ and $P^{(R)}$
 (cf.   eqs. (\ref{PLambda}), (\ref{PRJ}) and the Appendix) are found to
  be\footnote{We are choosing $\Omega^{23}=-\Omega^{32}=-1$.}
\begin{eqnarray}
P&=& \frac{1}{8} \frac{\left[(\pz)^{2}+(\pd)^{2}\right]}{\np^{6}}\label{P2}\\
P^{(R)}&=& -2 \left[2\sqrt{2}\frac{\pe}{\np^{2}}V_{0}V_{1}+\np^{2}(V_{1})^{2}\right]\label{PR2}.
\end{eqnarray}
For the functions $W_{\tx}$, $P_{\tx}$ and $P_{0}$ that enter the supersymmetry
transformation laws of the fermions, one obtains
\begin{eqnarray}
W_{1}&=&0\\
W_{2}&=&\frac{1}{4}\frac{\pd}{\np^{4}}\\
W_{3}&=&-\frac{1}{4}\frac{\pz}{\np^{4}},
\end{eqnarray}
respectively,
\begin{eqnarray}
P_{1}&=&\sqrt{2}\left(\sqrt{2}\frac{\pe}{\np^{4}}V_{0}-V_{1}\right)\\
P_{2}&=&-2\frac{\pz}{\np^{4}}V_{0}\\
P_{3}&=&-2\frac{\pd}{\np^{4}}V_{0}
\end{eqnarray}
and
\begin{equation}
P_{0}=\frac{2}{\sqrt{3}}\left(\frac{V_{0}}{\np^{2}}+\sqrt{2}\pe V_{1}\right).
\end{equation}

This shows that the necessary condition for an $\mathcal{N}=2$ supersymmetric
critical point, $W_{\tx}(\varphi_{c})=P_{\tx}(\varphi_{c})=0$, is equivalent to
\begin{eqnarray}
\langle \pz\rangle&=&\langle \pd\rangle=0\label{susy1}\\
\langle \pe\rangle^{3}V_{1}&=&\sqrt{2}V_{0}.\label{susy2}
\end{eqnarray}

Let us now analyze the critical points of the above scalar potentials. We will
 first investigate the critical points of $P$ and $P^{(R)}$ separately and then
consider the combined potential $P_{tot}=P+\lambda P^{(R)}$.\\

\noindent\textbf{\underline{The critical points of $\mathbf{P}$}}:\\
Taking the deriative of $P(\varphi)$ with respect to $\varphi^{\tx}$, one finds
\begin{eqnarray}
P_{,1}&=& -\frac{3}{4}\frac{\left[(\pz)^{2}+(\pd)^{2}\right]}{\np^{8}}\pe\\
&=&-A\pe + \frac{\pe}{4\np^{6}}\\
P_{,2}&=&A\pz\\
P_{,3}&=&A\pd,
\end{eqnarray}
where
\begin{displaymath}
A\equiv\frac{3}{4}\frac{\left[(\pz)^{2}+(\pd)^{2}\right]}{\np^{8}}+\frac{1}{4\np^{6}}
\end{displaymath}
has been introduced.
There are now two possibilities:\\
\underline{Case 1:} $A\neq 0$\\
Then $P_{,2}=P_{,3}=0$ implies $ \pz_{c}=\pd_{c}=0$ (which then also implies
$P_{,1}=0$). But then
$P(\varphi_{c})=0$,  and we have a Minkowski ground state, which, because of
$W_{\tx}(\varphi_{c})=0$, preserves  the full $\mathcal{N}=2$
supersymmetry (as long as the $U(1)_{R}$ gauging is turned off).\\
\underline{Case 2:} $A=0$\\
Then $P_{,1}=0$ implies $\frac{\pe}{4\np^{6}}=0$, which is inconsistent with
$\pe>0$ and $\np^{2}>0$.\\
\underline{Summary for $P$:}\\
 There exists a one parameter family of
$\mathcal{N}=2$ supersymmetric Minkowski ground states, given by
$\langle\pz\rangle=\langle\pd\rangle=0$ and arbitrary $\langle\pe\rangle>0$.
These vacua also preserve the $SO(2)$-gauge invariance.
There are no other critical points.\\

\noindent\textbf{\underline{The critical points of $\mathbf{P^{(R)}}$}:}\\
The gradient of $P^{(R)}$ is
\begin{eqnarray}
P^{(R)}_{,1}&=& -B\pe-\frac{4\sqrt{2}V_{0}V_{1}}{\np^{2}}\\
P^{(R)}_{,2}&=& B\pz\\
P^{(R)}_{,3}&=& B\pd,
\end{eqnarray}
where
\begin{displaymath}
B\equiv -8\sqrt{2}\frac{\pe}{\np^{4}}V_{0}V_{1} +4(V_{1})^{2}.
\end{displaymath}
There are now two possibilities:\\
\underline{Case 1:} $B=0$\\
$P^{(R)}_{,1}=0$ then requires $V_{0}V_{1}=0$. Thus either $V_{0}$ or $V_{1}$
(or both of them) have to
be zero. If $V_{0}=0$, $B=0$ implies $V_{1}=0$. Thus, $B=0$ automatically
implies $V_{1}=0$, and the potential $P^{(R)}$ vanishes identically  (cf.
    eq. (\ref{PR2})) resulting in a Minkowski vacuum. The $U(1)_{R}$-gauging
is non-trivial only when at least one $V_{I}$ is non-zero. Since $V_{1}=0$
in the case at hand, a non-trivial $U(1)_{R}$ gauging requires $V_{0}\neq 0$, implying
$P_{1}\neq 0$, ie. broken supersymmetry.\\
\underline{Case 2:} $B\neq 0$\\
The vanishing of $P^{(R)}_{,2}$ and $P^{(R)}_{,3}$ then requires
$\langle\pz\rangle=\langle\pd\rangle=0$, i.e.
 $\langle\np^{2}\rangle=\langle\pe\rangle^{2}$.
Because $P^{(R)}_{,1}$
has to vanish, this implies $\langle \pe\rangle^{3}(V_{1})^{2}=\sqrt{2}V_{0}V_{1}$.
 Thus there are two possibilities: Either $V_{1}=0$, or $V_{0}$ and $V_{1}$ are
 both non-vanishing. The former case leads us back to the case of identically
 vanishing potential $P^{(R)}\equiv 0$. The second possibility leads to a
 critical point with $\langle\pz\rangle=\langle\pd\rangle=0$ and
\begin{equation}
\langle \pe\rangle^{3}=\sqrt{2}\frac{V_{0}}{V_{1}}
\end{equation}
whenever $V_{0}V_{1}>0$ (since $\pe>0$). It is easy to see that this critical point
 satisfies the necessary conditions (\ref{susy1})-(\ref{susy2}) for $\mathcal{N}=2$
  supersymmetry.
The value of the potential $P^{(R)}$ at this  critical point is
\begin{equation}\label{cc}
P^{(R)}(\varphi_{c})=-6(\pe_{c})^{2}(V_{1})^{2}<0,
\end{equation}
i.e. it corresponds to an Anti-de Sitter ground state.\\
\underline{Summary for $P^{(R)}$:} \\There are three possibilities:\\
a) $V_{1}=0$.\\
This implies a flat potential $P^{(R)}\equiv 0$ and Minkowski ground states with
 broken supersymmetry (supersymmetry is broken as long as the $U(1)_{R}$-gauging
 is non-trivial, ie. if $V_{0}\neq 0$).\\
b) $V_{0}V_{1}>0$.\\
In this case, there exists exactly one critical point.
It is given by $\langle\pz\rangle=\langle\pd\rangle=0$ and
 $\langle\pe\rangle^{3}=\sqrt{2}V_{0}/V_{1}$
and corresponds to an $\mathcal{N}=2$ supersymmetric AdS ground state whose
cosmological constant can be read off from (\ref{cc}).
This vacuum breaks the global symmetry group $SO(1,1)\times SO(2,1)$ down to its
maximal compact subgroup  $SO(2)$.\\
c) $V_{0}V_{1}<0$.\\
No critical points exist in this case.\\

It is instructive to recover the characterization of the critical points
in terms of the dual element $V^{\#\ti}$ \cite{GST2} mentioned in section 3. Using
(\ref{Vdual}), one finds
\begin{displaymath}
V^{\#\ti}=((V_{1})^{2}/\sqrt{2},\sqrt{2}V_{0}V_{1},0,0).
\end{displaymath}
This shows that  $V^{\#\ti}=0$ is equivalent to $V_{1}=0$ and that $V^{\#\ti}$
is in the domain of positivity iff $V_{0}V_{1}>0$ so that our cases a), b), c)
are equivalent to the cases (i), (ii), (iii), respectively, listed in section 3.\\

\noindent\textbf{\underline{The critical points of the combined potential $\mathbf{P_{tot}=P+\lambda
 P^{(R)}}$}:}\\
The gradient of $P_{tot}$ is given by
\begin{eqnarray}
P_{tot , 1}&=& -(A+\lambda B)\pe +\frac{\pe}{4\np^{6}}-\lambda 4\sqrt{2}
\frac{V_{0}V_{1}}{\np^{2}}\\
P_{tot , 2}&=& (A+\lambda B)\pz\\
P_{tot , 3}&=& (A+\lambda B)\pd,
\end{eqnarray}
where $A$ and $B$ are again as defined above.

There are now two  possibilities:\\
\underline{Case 1:} $\langle\pz\rangle=\langle\pd\rangle=0$.\\
In this case, $P_{,\tx}(\varphi_{c})$ vanishes automatically (see the discussion
 of $P$ above). This implies that $P^{(R)}_{,\tx}(\varphi_{c})$ also has to
  vanish separately, i.e. we are dealing with critical points that are just
   simultaneous critical points of the individual potentials $P$ and $P^{(R)}$.
    These have already been discussed above.\\
\underline{Case 2:} $\langle\pz\rangle^{2}+\langle\pd\rangle^{2}>0$.\\
This case involves a nontrivial interplay of the two potentials $P$ and
$P^{(R)}$. For $P_{tot ,2}$ and $P_{tot ,3}$ to vanish, one obviously needs
$A+\lambda B=0$. $P_{tot , 1}=0$ then implies
\begin{equation}\label{cond1}
\frac{\pe}{\np^{4}}=16\sqrt{2}\lambda V_{0}V_{1}.
\end{equation}
This implies (remembering $\lambda>0$ and $\pe>0$)
\begin{equation}\label{cond1b}
V_{0}V_{1}> 0.
\end{equation}
Inserting (\ref{cond1}) into $A+\lambda B=0$, and reexpressing $(\pz)^{2}+
(\pd)^{2}$ in terms of ${\np}^{2}$ and $(\pe)^{2}$, one derives the
additional condition
\begin{equation}\label{cond2}
\frac{1}{\np^{6}}=\frac{1}{2}(16\sqrt{2}\lambda V_{0}V_{1})^{2} +8\lambda (V_{1})^{2}.
\end{equation}
Now, by assumption,   $\langle\pz\rangle^{2}+\langle\pd\rangle^{2}>0$. Hence
\begin{displaymath}
\frac{(\pe)^{2}}{\np^{8}}> \frac{1}{\np^{6}},
\end{displaymath}
so that in order for (\ref{cond1}) and (\ref{cond2}) to be consistent,
one needs
\begin{equation}\label{cond3}
32\lambda(V_{0})^{2}>1.
\end{equation}
Thus, if $V_{0}$ is  big enough such that (\ref{cond3}) is fulfilled and if
$V_{1}V_{0}>0$ (cf. (\ref{cond1b})), new non-trivial critical points exist.
Eq. (\ref{cond2}) fixes $\|\varphi_{c}\|^{2}$ so that eq. (\ref{cond1}) fixes
$\pe_{c}$. This in turn fixes $((\pz_{c})^{2}+(\pd_{c})^{2})$, but not
$\pz_{c}$ and $\pd_{c}$ individually. Hence, we obtain a one-parameter family
of critical points, which, because of $((\pz_{c})^{2}+(\pd_{c})^{2})>0$,
do not preserve the full $\mathcal{N}=2$ supersymmetry  (cf. (\ref{susy1}))
 and spontaneously break
 the $SO(2)$-gauge invariance. Using (\ref{cond1}) and (\ref{cond2}), one
finds for the value of $P_{tot}$ at these critical points
\begin{equation}
P_{tot}(\varphi_{c})=-\frac{3}{8}\frac{1}{\np^{4}}<0,
\end{equation}
which again corresponds to an Anti-de Sitter solution. Putting everything together,
 we arrive at the following\\
\underline{Summary for $P_{tot}$:}\\
Depending on the values of the $V_{I}$, the total potential $P_{tot}=P+\lambda P^{(R)}$
 admits the following types of critical points:\\
a) $V_{1}=0$. \\ In this case, $P^{(R)}$ vanishes identically, and one has a
one-parameter family of $SO(2)$ gauge invariant Minkowski ground states.
They are given by $\pz_{c}=\pd_{c}=0$ and an arbitrary $\pe_{c}>0$. If $V_{0}\neq 0$
 (i.e. if the $U(1)_{R}$-gauging is non-trivial), these ground states
break the $\mathcal{N}=2$ supersymmetry. If $V_{0}=0$, the $U(1)_{R}$-gauging
is switched off, and supersymmetry is unbroken, corresponding to  case 1
in the discussion of $P$.\\
b1) $V_{0}V_{1}>0$, and $32\lambda (V_{0})^{2} \leq 1$.\\
In this case, there is precisely one ground state. It preserves the full $\mathcal{N}=2$
 supersymmetry and the $SO(2)$ gauge invariance. It corresponds to an Anti-de Sitter
 solution, and is given by $\pz_{c}=\pd_{c}=0$
and $(\pe_{c})^{3}=\sqrt{2}V_{0}/V_{1}$ with
 $P_{tot}(\varphi_{c})=\lambda P^{(R)}(\varphi_{c})= -6\lambda(\pe_{c})^{2}(V_{1})^{2}$.
 Although the potential $P$ due to the tensor fields does not contribute to this
 cosmological constant,
it does have an   effect on the form of the extremum  of the total potential: It
 is now a saddle point, as opposed to the case of pure $U(1)_{R}$ gauging, where
  the supersymmetric critical point is always a maximum.\\
b2) $V_{0}V_{1}>0$, and in addition $32\lambda (V_{0})^{2}>1$\\
In this case, there are two types of critical points. The first one is an
isolated supersymmetric critical point which has exactly the same properties as
 the one described in b1) above, with one exception: it is now a local
  \emph{maximum} of the total scalar potential.
Apart from this point, there is an additional one-parameter family of
critical points. They are given by eqs. (\ref{cond1}) and (\ref{cond2}), which
 fix $\pe_{c}$, and $[(\pz_{c})^2+(\pd_{c})^2]$. They break the $\mathcal{N}=2$
supersymmetry and the $SO(2)$-gauge invariance and correspond to an
 Anti-de Sitter solution with $P_{tot}(\varphi_{c})=-3/(8\|\varphi_{c}\|^{4})$.
These critical points are saddle points of the total potential.\\
c)  $V_{0}V_{1}<0$. \\
In this case, there are no critical points.\\

\subsection{The $U(1)_{R}\times SO(1,1)$ gauging}
We now come to the noncompact version of the above theory. Since the analysis
is very similar to the compact case, our presentation can be less detailed.

We choose the
$SO(1,1)$ subgroup of $SO(2,1)$ to rotate  the components $\xi^{1}$ and $\xi^{2}$ into
 each other.
Consequently, this  $SO(1,1)$ acts nontrivially on the vector fields $A_{\mu}^{1}$ and
 $A_{\mu}^{2}$, and its gauging requires the dualization of
$A_{\mu}^{1}$ and $A_{\mu}^{2}$ to antisymmetric tensor fields. Accordingly,
we decompose the index $\ti$ as follows
\begin{displaymath}
\ti=(I,M)
\end{displaymath}
with $I,J,K,\ldots=0,3$ and $M,N,P,\ldots=1,2$.

Since $C_{0MN}\neq 0$ and $C_{3MN}=0$,
$A_{\mu}^{0}$ plays the r\^{o}le of the $SO(1,1)$-gauge field, whereas
 $A_{\mu}^{3}$ is  a ``spectator vector field'' with respect to  the
$SO(1,1)$-gauging.

Using a linear combination $A_{\mu}[U(1)_{R}]=A_{\mu}^{I}V_{I}$ of the
vector fields $A_{\mu}^{0}$ and $A_{\mu}^{3}$ as the $U(1)_{R}$-gauge field,
one can then simultaneously gauge $U(1)_{R}$ and $SO(1,1)$, and obtains the
 ($U(1)_{R}\times SO(1,1)$)-gauged analog of the ($U(1)_{R}\times SO(2)$)-theory
 discussed in the previous subsection.

The scalar potentials $P$ and $P^{(R)}$ are now (we use $\Omega^{12}=-\Omega^{21}=-1$)
\begin{eqnarray}
P&=& \frac{1}{8} \frac{\left[(\pe)^{2}-(\pz)^{2}\right]}{\np^{6}}\label{P3}\\
P^{(R)}&=& -2 \left[2\sqrt{2}\frac{\pd}{\np^{2}}V_{0}V_{3}-\np^{2}(V_{3})^{2}\right]\label{PR3}.
\end{eqnarray}
For the functions $W_{\tx}$, $P_{\tx}$ and $P_{0}$ that enter the supersymmetry
transformation laws of the fermions, one obtains
\begin{eqnarray}
W_{1}&=&-\frac{1}{4}\frac{\pz}{\np^{4}}\\
W_{2}&=&\frac{1}{4}\frac{\pe}{\np^{4}}\\
W_{3}&=&0,
\end{eqnarray}
respectively,
\begin{eqnarray}
P_{1}&=&2\frac{\pe}{\np^{4}}V_{0}\\
P_{2}&=&-2\frac{\pz}{\np^{4}}V_{0}\\
P_{3}&=&-\sqrt{2}\left(\sqrt{2}\frac{\pd}{\np^{4}}V_{0}+V_{3}\right)
\end{eqnarray}
and
\begin{equation}
P_{0}=\frac{2}{\sqrt{3}}\left(\frac{V_{0}}{\np^{2}}+\sqrt{2}\pd V_{3}\right).
\end{equation}

This already shows that there can be  no $\mathcal{N}=2$ supersymmetric critical point,
 because $W_{2}$ can never vanish.

Let us now come to the critical points of the scalar potentials. We will again first
 analyse the critical points of $P$ and $P^{(R)}$ separately and then
consider the combined potential $P_{tot}=P+\lambda P^{(R)}$.\\

\noindent\textbf{\underline{The critical points of $\mathbf{P}$}}:\\
For the gradient of $P(\varphi)$ with respect to $\varphi^{\tx}$, one obtains
\begin{eqnarray}
P_{,1}&=& \tilde{A}\pe\\
P_{,2}&=&-\tilde{A}\pz\\
P_{,3}&=&-\tilde{A}\pd + \frac{\pd}{4\np^{6}}
\end{eqnarray}
with
\begin{displaymath}
\tilde{A}\equiv-\frac{3}{4}\frac{\left[(\pe)^{2}-(\pz)^{2}\right]}{\np^{8}}+
\frac{1}{4\np^{6}}.
\end{displaymath}
Since $\varphi^{1}$ cannot vanish, $P_{,1}=0$ requires $\tilde{A}=0$\\
But then $P_{,3}=0$ implies $\pd=0$. The assumption $\tilde{A}=0$ then leads to
the contradiction $1=3$, and is therefore inconsistent.\\
\underline{Summary for $P$:}\\
$P$ alone has no critical points at all.\\

\noindent\textbf{\underline{The critical points of $\mathbf{P^{(R)}}$}:}\\
The gradient of $P^{(R)}$ is
\begin{eqnarray}
P^{(R)}_{,1}&=& \tilde{B}\pe\\
P^{(R)}_{,2}&=& -\tilde{B}\pz\\
P^{(R)}_{,3}&=& -\tilde{B}\pd-\frac{4\sqrt{2}V_{0}V_{3}}{\np^{2}},
\end{eqnarray}
where
\begin{displaymath}
\tilde{B}\equiv 8\sqrt{2}\frac{\pd}{\np^{4}}V_{0}V_{3} +4(V_{3})^{2}.
\end{displaymath}
Since $\pe$ cannot vanish, $P^{(R)}_{,1}=0$ implies $\tilde{B}=0$. The condition
 $P^{(R)}_{,3}=0$ then implies $V_{0}V_{3}=0$. Assume $V_{3}\neq 0$. Then $V_{0}=0$
  would imply $V_{3}= 0$ by virtue of $\tilde{B}=0$. Thus, $V_{3}$ has to vanish in
  any case if a critical point of $P^{(R)}$ is assumed to exist. However,  $P^{(R)}$
  then vanishes identically.\\
\underline{Summary for $P^{(R)}$:}\\
A critical point of $P^{(R)}$ exists if and only if $P^{(R)}$ vanishes identically
 (which is equivalent to $V_{3}=0$).\\

It is easy to recover the characterization of the critical points of $P^{(R)}$
in terms of the dual element $V^{\#\ti}$ \cite{GST2} mentioned in section 3.
In the case at hand, one finds
\begin{displaymath}
V^{\#\ti}=(-(V_{3})^{2}/\sqrt{2},0,0,-\sqrt{2}V_{0}V_{3}).
\end{displaymath}
This shows that  $V^{\#\ti}=0$ is equivalent to $V_{3}=0$ and that $V^{\#\ti}$
can never be in the domain of positivity if $V_{3}\neq 0$. Thus, our results are
 consistent with the discussion given in section 3.\\

\noindent\textbf{\underline{The critical points of the combined potential
 $\mathbf{P_{tot}=P+\lambda P^{(R)}}$:}}\\
The gradient of $P_{tot}$ is given by
\begin{eqnarray}
P_{tot , 1}&=& (\tilde{A}+\lambda \tilde{B})\pe\\
P_{tot , 2}&=& -(\tilde{A}+\lambda \tilde{B})\pz\\
P_{tot , 3}&=& -(\tilde{A}+\lambda \tilde{B})\pd+\frac{\pd}{4\np^{6}}-\lambda 4\sqrt{2}
\frac{V_{0}V_{3}}{\np^{2}},
\end{eqnarray}
where $\tilde{A}$ and $\tilde{B}$ are again as defined above.

Since $\pe$ cannot vanish, $P_{tot , 1}=0$ requires $(\tilde{A}+\lambda \tilde{B})=0$.
 $P_{tot , 3}=0$ then implies
\begin{equation}\label{cond4}
\frac{\pd}{\np^{4}}=16\sqrt{2}\lambda V_{0}V_{3}.
\end{equation}
The analogous equation (\ref{cond1}) in the compact gauging implied $V_{0}V_{1}>0$
(cf. eq. (\ref{cond1b})). In the case at hand, however, eq. (\ref{cond4})
does not imply any constraint for $V_{0}V_{3}$, because $\pd/\np^{4}$ does not have to
 be positive, as opposed to $\pe/\np^{4}$, which is always greater than zero.

Inserting (\ref{cond4}) into $\tilde{A}+\lambda \tilde{B}=0$, one derives the
additional condition (i.e. the analog of (\ref{cond2}))
\begin{equation}\label{cond5}
\frac{1}{\np^{6}}=-\frac{1}{2}(16\sqrt{2}\lambda V_{0}V_{3})^{2} +8\lambda (V_{3})^{2}.
\end{equation}

Since $1/\np^{6}>0$, the last equation implies the consistency conditions
\begin{eqnarray}
V_{3}&\neq& 0\nonumber\\
32\lambda(V_{0})^{2}&<&1. \label{cond6}
\end{eqnarray}
(The analogous equation (\ref{cond3}) in the compact gauging arose
as a consistency condition of (\ref{cond1}) and (\ref{cond2}). However, it is easy to
see that (\ref{cond4}) and (\ref{cond5}) do not imply any additional constraints on
$V_{0}$ or $V_{3}$, so that eqs. (\ref{cond6}) remain the only constraints on the $V_{I}$.)

For a given set of $V_{I}$ and $\lambda$ subject to (\ref{cond6}), $\np^{2}$
is fixed by (\ref{cond5}). This in turn fixes $\pd$ and $((\pe)^{2}-(\pz)^{2})$
 by virtue of eq. (\ref{cond4}), but leaves the $\pe$ and $\pz$ otherwise
 undetermined. We thus obtain a one parameter family of critical points
which can be viewed as the noncompact analog of the nontrivial non-supersymmetric
critical points found for the compact gauging (i.e. the ones mentioned in case
 b2) in the discussion of $P_{tot}$). However, for the non-compact
gauging, these critical points have very different physical properties.
In particular, the total scalar potential becomes
\begin{equation}
P_{tot}(\varphi_{c})=3\lambda \np^{2}(V_{3})^{2}[1-32\lambda (V_{0})^{2}],
\end{equation}
which is positive because of the condition (\ref{cond6}) and therefore
corresponds to a \emph{de Sitter} rather than an Anti-de Sitter spacetime.\\
\underline{Summary for $P_{tot}$:}\\
If $V_{3}\neq 0$ and $32\lambda(V_{0})^{2}<1$, there exists a one parameter family
 of critical points given by (\ref{cond4}) and (\ref{cond5}). They correspond to a de
  Sitter spacetime with $P_{tot}(\varphi_{c})=3\lambda \np^{2}(V_{3})^{2}[1-32\lambda
  (V_{0})^{2}]>0$ and break the $\mathcal{N}=2$ supersymmetry and the $SO(1,1)$ gauge
  invariance. There are no other critical points of the combined potential.
In particular, neither the analog of the $\mathcal{N}=2$ supersymmetric critical point
 mentioned in case b1) and b2), nor the analogs of the
Minkowski ground states mentioned in case a) in the summary for $P_{tot}$
in section 4.2,
 exist for the non-compact gauging.

\section{The generic Jordan family of $\mathcal{N}=2$ gauged Yang-Mills/Einstein
supergravity theories coupled to tensor multiplets}
In the previous section we studied in detail the critical points of
the potentials of the simplest non-trivial gauged Yang-Mills Einstein
supergravity theories with tensor multiplets. The corresponding
$\mathcal{N}=2$ MESGT belongs to the generic Jordan family and has the scalar
manifold $SO(1,1)\times SO(2,1)/SO(2)$. The
MESGT's of the generic Jordan family  have the scalar manifold
 $SO(1,1)\times SO(n-1,1)/SO(n-1)$.
From the results of  \cite{gz99} and the arguments given in the previous section
it follows that any gaugeable     subgroup $K$ of the
isometry group with $K$-charged vectors dualized to tensor fields must
be Abelian. Since the vector field $A^0_{\mu}$ must be the gauge
field it follows that one can only gauge $SO(2)$ or $SO(1,1)$ and
have some $K$-charged tensor fields under them. We should also note that the
gaugeable     $SO(1,1)$ must be a subgroup of $SO(n-1,1)$ and can not be
the $SO(1,1)$ factor in the isometry group since all the vector
fields are charged under the latter $SO(1,1)$. The  $SO(2)$
gauge group is some diagonal subgroup of the maximal Abelian subgroup
$SO(2)_1\times SO(2)_2 \times \cdots \times SO(2)_p$ of $SO(n-1,1)$
(for $n=2p+1$ or $n=2p+2$). The gaugeable     $SO(1,1)$  subgroup is
unique modulo some $SO(n-1)$ rotation.

 After the gauging of the
Abelian subgroup of the isometry group with the charged vectors
dualized to tensor fields, the remaining vector fields can be used to
gauge some non-Abelian subgroup $S$ of the full isometry group so long as
they decompose as the adjoint plus some singlets of $S$.
This non-Abelian gauging does not introduce any additional potential
\cite{GST2}. A linear combination of the remaining $S$ singlet vector
fields can then be used to gauge the $U(1)_R$ subgroup of the
$R$-symmetry group $SU(2)_{R}$.
The full potential of the $K\times U(1)_R\times S$ gauged Yang-Mills
Einstein supergravity with $K$-charged tensor fields  must have novel
 critical points of the type we discussed in
the previous section since these theories can be truncated to the the
simplest non-trivial model consistently.

There exist an infinite family of non-Jordan MESGT's with the scalar
 manifold $SO(n,1)/SO(n)$ \cite{GST3}. For this family only
the parabolic  subgroup $SO(n-1)\times SO(1,1)\odot T_{(n-1)}$,
which is simply an ``internal Euclidean group''
 in $(n-1)$ dimensions times a dilatation
 factor,
extends to a symmetry of the full action \cite{dWvP2}.
 The analysis of the possible
gauge groups $K$ that involve a dualization of $K$-charged vectors to tensor fields
is very similar to the generic Jordan case \cite{gz99}. In this case too one
finds that only a one dimensional Abelian subgroup $K$ can be gauged
with nontrivial tensor fields carrying charge under $K$.
However, there is one crucial difference between the Jordan family
and the non-Jordan family. For the non-Jordan family the tensor $C_{\ti\tj\tk}$ is
 not an invariant tensor of the full isometry group $SO(n,1)$ of the scalar
manifold. As a consequence
one finds that
\begin{equation}
  C_{\ti\tj\tk}\neq C^{\ti\tj\tk},
\end{equation}
and the  $C^{\ti\tj\tk}$ are no longer constant tensors but depend on
the scalars.

\renewcommand{\theequation}{\arabic{section}.\arabic{equation}}

\section{Conclusions}

In this paper we have analyzed the scalar potentials of the simplest examples
of a gauged Yang-Mills/Einstein supergravity theory coupled to tensor multiplets.

Although not all the results we have derived for these
 examples may carry over to the most general gauged Yang-Mills/Einstein
 supergravity theory with tensor fields, they show that the scalar potentials
 of these theories can exhibit  a much richer structure than the purely
  $U(1)_{R}$-gauged supergravity theories or the gauged
 Yang-Mills/Einstein    supergravity theories \emph{without} tensor fields.
Our analysis revealed   that even though the total potential is just a sum of
the potentials that appear in the separate gaugings of $K$ and $U(1)_{R}$,
there can  be critical points of the total potential which would not be critical
 points of the individual potentials.
In particular, we found that, for a certain parameter range (case b2) in Section 4.2),
 the ($U(1)_{R}\times SO(2))$ gauging leads to a new one-parameter family of
  non-supersymmetric critical points, which are saddle points of the total
potential. These are accompanied by an isolated $\mathcal{N}=2$ supersymmetric
\emph{maximum}, which is already present in the purely $U(1)_{R}$ gauged theory without
 tensor fields. In another parameter range (case b1)), the novel non-supersymmetric
  one-parameter family of critical points disappears and  the $\mathcal{N}=2$
   supersymmetric critical point becomes a
\emph{saddle point} (and remains supersymmetric). In yet another parameter range
(case a)), the theory has a one-parameter family of Minkowski ground states
which break the $\mathcal{N}=2$ supersymmetry as long as the $U(1)_{R}$ gauging is
 nontrivial. If the $U(1)_{R}$ gauging is switched off, these critical points become
  supersymmetric.

The possible types of critical points
are much more restricted for the non-compact $U(1)_{R}\times SO(1,1)$ gauging,
which can have at most a one-parameter family of non-supersymmetric \emph{de Sitter}
ground states (which are presumably unstable).
 This is consistent
with the experience from compact and non-compact gaugings of the
$\mathcal{N}=8$ theory \cite{GRW1} where a non-supersymmetric
de Sitter critical point was found in the $SO(3,3)$-gauged version of
the $\mathcal{N}=8$ theory.

In this paper we have not studied the critical points of the
potential when one gauges a \emph{non-Abelian} subgroup $K$ of the isometry
group of the scalar manifold with tensor multiplets transforming in a
nontrivial representation of $K$. Such gauge groups are possible for the
magical Jordan $\mathcal{N}=2$ theories as well as for the infinite family of theories
with $SU(n)$ isometries discussed in \cite{gz99}. The study of the
critical points of these theories as well as those of the non-Jordan
family  will be the subject of a future investigation.

\appendix
\section{The ``very special geometry'' of the $SO(1,1)\times SO(2,1)/SO(2)$-model}
This Appendix contains a list of the basic scalar field dependent quantities
that enter the Lagrangian
and the transformation laws of the ungauged and gauged
 $SO(1,1)\times SO(2,1)/SO(2)$-theory.

In our parametrization, the $h^{\ti}=\sqrt{\frac{2}{3}}\xi^{\ti}|_{N=1}$
are
\begin{displaymath}
h^{0}=\frac{1}{\sqrt{3}\np^{2}}, \qquad
h^{1}=\sqrt{\frac{2}{3}}\pe,  \qquad
h^{2}=\sqrt{\frac{2}{3}}\pz,  \qquad
h^{3}=\sqrt{\frac{2}{3}}\pd.
\end{displaymath}
For the $h_{\ti}=\frac{1}{\sqrt{6}}\frac{\partial}{\partial \xi^{\ti}}N|_{N=1}$
one obtains
\begin{displaymath}
h_{0}=\frac{1}{\sqrt{3}}\np^{2}, \qquad
h_{1}=\frac{2}{\sqrt{6}}\frac{\pe}{\np^{2}}, \qquad
h_{2}=-\frac{2}{\sqrt{6}}\frac{\pz}{\np^{2}}, \qquad
h_{3}=-\frac{2}{\sqrt{6}}\frac{\pd}{\np^{2}}.\nonumber
\end{displaymath}
The vector/tensor field metric ${\stackrel{\scriptscriptstyle{\circ}}{a}}_{\ti\tj}
=-\frac{1}{2}\frac{\partial}{\partial\xi^{\ti}}\frac{\partial}{\partial\xi^{\tj}}
\ln N(\xi)|_{N=1}$ turns out to be
\begin{displaymath}
{\stackrel{\scriptscriptstyle{\circ}}{a}}_{\ti\tj}=\left(
\begin{array}{cccc}
\np^{4}&0&0&0\\
0&2(\pe)^{2}\np^{-4}-\np^{-2}&-2\pe\pz\np^{-4}&-2\pe\pd\np^{-4}\\
0&-2\pe\pz\np^{-4}&2(\pz)^{2}\np^{-4}+\np^{-2}&2\pz\pd\np^{-4}\\
0&-2\pe\pd\np^{-4}&2\pz\pd\np^{-4}&2(\pd)^{2}\np^{-4}+\np^{-2}
\end{array}
\right).
\end{displaymath}
This shows that the unique point with ${\stackrel{\scriptscriptstyle{\circ}}{a}}_{\ti\tj}
=\delta_{\ti\tj}$ corresponds to $\varphi^{\tx}=(1,0,0)$, as has been mentioned
earlier.\footnote{If we had chosen another normalization for $N$, ie. $N(\xi)=a\xi^{0}
\left[(\xi^{1})^{2}-(\xi^{2})^{2}-(\xi^{3})^{2}\right]$ for some $a\in\Rbar$,
${\stackrel{\scriptscriptstyle{\circ}}{a}}_{00}$ would have been $a^{2}\np^{4}/2$ with
 the other components unchanged. It is easy to see that only $a=\sqrt{2}$ can lead to
a point where ${\stackrel{\scriptscriptstyle{\circ}}{a}}_{\ti\tj}=\delta_{\ti\tj}$.}

Finally, the metric $g_{\tx\ty}$ on $\mathcal{M}$ reads
\begin{displaymath}
g_{\tx\ty}=\left(
\begin{array}{ccc}
4(\pe)^{2}\np^{-4}-\np^{-2}&-4\pe\pz\np^{-4}&-4\pe\pd\np^{-4}\\
-4\pe\pz\np^{-4}&4(\pz)^{2}\np^{-4}+\np^{-2}&4\pz\pd\np^{-4}\\
-4\pe\pd\np^{-4}&4\pz\pd\np^{-4}&4(\pd)^{2}\np^{-4}+\np^{-2}
\end{array}
\right).
\end{displaymath}
For the determinants of ${\stackrel{\scriptscriptstyle{\circ}}{a}}_{\ti\tj}$ and
 $g_{\tx\ty}$,
one finds
\begin{eqnarray}
\det {\stackrel{\scriptscriptstyle{\circ}}{a}}_{\ti\tj}&=&\np^{-2}\\
\det g_{\tx\ty}&=&3\np^{-6},
\end{eqnarray}
which shows that ${\stackrel{\scriptscriptstyle{\circ}}{a}}_{\ti\tj}$
 and $g_{\tx\ty}$ are
positive definite and well-behaved throughout the entire ``positive timelike''
 region (i) and
that both are
not positive definite in region (ii), where $\np^{2}<0$.

{\bf Acknowledgements:} We would like to thank Eric Bergshoeff, Renata Kallosh,
Andrei Linde and Toine van Proeyen for fruitful discussions.

\end{document}